\documentstyle[prb,aps,preprint]{revtex}

\begin{document}
\title{Linear spin and orbital wave theory for undoped manganites}
\author{R. Y. Gu, Shun-Qing Shen, Z. D. Wang}
\address{Department of Physics,The University of Hong Kong, Pokfulam, Hong Kong, China}
\author{D. Y. Xing}
\address{National Center for Theoretical Sciences, P.O.Box 2-131, 
Taiwan 300,
and National Laboratory of Solid State Microstructures, Nanjing
University, Nanjing 210093, China }

\date{\today}
\maketitle

\begin{abstract}
We present a linear spin and orbital wave theory to account for the
spin and orbital orderings observed experimentally in undoped manganite. 
It is found that the anisotropy of the magnetic structure is closely
related to the orbital ordering, and the Jahn-Teller
effect stabilizes the orbital ordering. The phase diagram
and the low energy excitation spectra for both spin and orbital orderings 
are obtained. The calculated critical temperatures can be 
quantitatively comparable to the experimental data.
\end{abstract}

\pacs{PACS numbers: 71.45.Lr, 75.30.Mb, 75.10-b}
\newpage

LaMnO$_{3}$ is the parent compound of colossal magnetoresistance (CMR)
manganites, and has been studied both experimentally and theoretically.
The compound is an insulator with layered antiferromagnetic (A-type AF)
spin ordering and an orbital ordering of e$_{g}$ electrons.
\cite{goodenough} Murakami {\it et al}.\cite{murakami} has recently
succeeded in detecting the orbital ordering in LaMnO$_{3}$ by using
resonant x-ray scattering techniques with the incident photon energy
tuned near the Mn K-absorption edge. The orbital order parameter
decreases above the Neel
temperature T$_{N}\sim 140K$ and persists until T$_{O}\sim 780K.$ 
Theoretically, the problem of orbital degeneracy in a d-electron system
was pioneered by Kugel and Khomskii \cite{kugel} in 1970's and \
investigated extensively in recent years.
\cite{ishihara1,moreo,shiina,maezono,brink,hotta,millis1,shen,feiner}

In this paper, starting from an effective Hamiltonian of
the spin and orbital interactions, as well as the JT coupling
between the $e_g$ electrons and the lattice distortion,
we investigate the interplay among the spin, orbit and the lattice
distortion. We present the phase diagram as functions of interaction
parameters, and obtain the low-energy excitations of the system in
different phases. It is found that special properties of the orbital
operators can result in an anisotropy of the magnetic
structure and an energy gap of the orbital excitations.
We also estimate the critical temperatures for spin and orbital
orderings as well as their dependence on the JT coupling. The calculated
results are comparable to the experimental measurements.

The effective spin and orbital interactions are derived by the
projection perturbation method up to the second order
\cite{shiina,shen,feiner}
\begin{eqnarray}
H_{e}^{eff} &=&J_{1}\sum_{ij}({\bf S}%
_{i}\cdot {\bf S}_{j}-4)n_i^{\alpha }n_j^{\alpha } 
+J_{2}\sum_{ij}({\bf S}_{i}\cdot 
{\bf S}_{j}-4)n_i^{\alpha }n_j^{\bar{\alpha}} \nonumber \\
&-&J_{3}\sum_{ij}[{\bf S}_{i}\cdot 
{\bf S}_{j}+6]n_i^{\alpha }n_j^{\bar{\alpha}}\ ,
\label{heff}
\end{eqnarray}
where ${\bf S}_{i}$ is the spin operator of $S=2$.
The three terms describe three processes with different intermediate
states.\cite{shen,strength}
$n_i^{\alpha }=d_{i\alpha}^{\dag }d_{i\alpha }$ and $n_i^{\bar{\alpha}}=
d_{i\bar{\alpha}}^{\dag }d_{i \bar{\alpha}}$ are the particle number
operators of $e_{g}$ electron in orbit states $|\alpha \rangle =
\cos (\varphi _{\alpha}/2)|z\rangle +\sin (\varphi_{\alpha }/2)
|\bar{z}\rangle $ and $|\bar{\alpha}\rangle =-\sin (\varphi _{\alpha
}/2)|z\rangle +\cos (\varphi _{\alpha}/2)|\bar{z}\rangle $, respectively,
with orbital states $|z\rangle \propto (3z^{2}-r^{2})/\sqrt{3}$ and
$|\bar{z}\rangle \propto x^{2}-y^{2}$. Here $\varphi_{\alpha}$ depends on
the direction of the $(ij)$ bond by  $\varphi _{x}=-2\pi /3$, $\varphi
_{y}=2\pi /3$, and $\varphi _{z}=0$, respectively, for bond $(ij)$
parallel to the $x$, $y$ and $z$ directions. 
The introduced $d_{i\alpha }^{\dag },d_{i\alpha }$ and $d_{i%
\bar{\alpha}}^{\dag },d_{i\bar{\alpha}}$ are operators in the orbital space,
with $d_{i\alpha }^{\dag }|0\rangle =|\alpha \rangle $, $d_{i\bar{\alpha}}^{\dag }|0\rangle =|%
\bar{\alpha}\rangle $, they should satisfy the constraint $n_i^{\alpha }+n_i^{\bar{%
\alpha}}=1$.

The JT interaction can be expressed as \cite{kanamori} 
\begin{equation}
H_{JT}=-g\sum_{i\gamma\gamma^{\prime}}
d^{\dag}_{i\gamma}{\bf T}_{\gamma\gamma^{\prime}}\cdot {\bf Q}_id_{i\gamma^{\prime}}
+\frac{K}{2}\sum_i{\bf Q}_i^2\ ,
\end{equation}
where ${\bf T}=(T_x,T_z)$ are the Pauli matrices in the orbital space with
$\gamma$ $ (\gamma^{\prime})=z$ or $\bar{z}$, and $g$ is the
coupling between the $e_g$ electrons and the local JT lattice distortion $%
{\bf Q}_i=Q_i(\sin\phi_i, \cos\phi_i)$. Here we have neglected the terms
for the anharmonic oscillation of JT distortion and the higher order
coupling, their effect being regarded approximately as giving a preferable
direction $\phi _{i}$ of the JT distortion observed experimentally.

Experimental measurement on LaMnO$_3$ indicates that the critical
temperature of the
orbital ordering, $T_{O}$, is much higher than that of the magnetic
ordering, $T_{N}$, \cite{murakami} As a result, the spin and orbital
degrees of freedom, which are coupled to each other in Hamiltonian
(\ref{heff}), may be separately treated by the Hartree-Fock mean-field
approach. The total Hamiltonian is reduced to 
$H_{MF}=H_{S}+H_{O}+E_{0}$,
where $H_{S}$ and $H_{O}$ are the decoupled spin and orbital Hamiltonians,
respectively, and $E_{0}$ is an energy constant. The spin
Hamiltonian $H_{S}$ is given by 
\begin{equation}
H_{S}=\sum_{(ij)}\tilde{J}_{ij}{\bf S}_{i}\cdot {\bf S}_{j}\
,  \label{spin}
\end{equation}
with the effective spin coupling depending on
the orbital configuration of the two neighboring sites by 
\begin{eqnarray}
\tilde{J}_{ij} &=&\frac{1}{2}J_1\langle (1+m_i^{\alpha })(1+m_j^{\alpha })%
\rangle 
\nonumber \\
&\ &+\frac{1}{2}%
(J_{2}-J_{3})\langle 1-m_i^{\alpha}m_j^{\alpha}\rangle +\tilde{J}_{AF}\ ,
\end{eqnarray}
where  $m_i^{\alpha }=n_i^{\alpha }-n_i^{\bar{\alpha}}$ are the orbital
operators, and the $\tilde{J}_{AF}$ term comes from the magnetic
superexchange between the nearest neighboring local spins. It is worthy
of pointing out here that the orbital operators introduced above has
unusual operator algebra, quite different from that of the spin operators.
It can be shown that they satisfy the following relations:
$(m_i^{\alpha})^2=1, m_i^x+m_i^y+m_i^z=0,$ and
$[m_i^x, m_i^y]=[m_i^y, m_i^z]=[m_i^z, m_i^x]=%
\sqrt{3}(d^{\dag}_{i\bar{z}}d_{iz}-d^{\dag}_{iz}d_{i\bar{z}})$.
The orbital Hamiltonian $H_O$ can be written as
\begin{eqnarray}
H_{O} &=&\sum_{(ij)}u_{ij}m_i^{\alpha }m_j^{\alpha
}-\sum_{(ij)}h_{ij}m_i^{\alpha }+\frac{K}{2}\sum_{i}Q_{i}^{2} \nonumber \\
&&-g\sum_{i}Q_{i}\left( m_i^z\cos \phi _{i}+\frac{1}{\sqrt{3}}%
(m_i^y-m_i^x)\sin \phi _{i}\right) \ ,
\end{eqnarray}
where the effective orbital coupling $u_{ij}$ depends on the
spin configuration of the two neighboring sites by 
\begin{eqnarray*}
u_{ij} &=&\frac{1}{2}(J_{1}-J_{2}+J_3)\langle {\bf S}_{i}\cdot {\bf S}%
_{j}\rangle
+(3J_3-2J_1+2J_2)\ ,
\end{eqnarray*}
and  
$
h_{ij}=-\frac{1}{2}J_{1}\langle{\bf S}_{i}\cdot {\bf S}_{j}-4
\rangle \ .
$ All these coupling parameters $\tilde{J}_{ij}$, $u_{ij}$ and $h_{ij}$ in
$H_S$ and $H_O$ are determined not only by the spin and orbital
configurations of the nearest neighboring sites $i$ and $j$, but also by
the direction of the (ij) bond. For short, we denote them by $\tilde{J}%
_{\alpha}$, $u_{\alpha}$ and $h_{\alpha}$ thereafter. If there are two
symmetric directions in the system, e.g., $x$- and $y$-direction,
one has $\tilde{J}_x=\tilde{J}_y, u_x=u_y$, and $h_x=h_y$.

The spin Hamiltonian $H_{S}$ is an
anisotropic Heisenberg Hamiltonian with SU(2) symmetry. At low
temperatures, the spin configuration along the $\alpha $ direction is
determined by the sign of $J_{\alpha }$. Dividing the system into two 
sublattices $A$ and $B$ according to their spin alignments, and
performing the well-known Holsten-Primakoff (HP) transformation in the
linear spin wave (LSW) theory, up to the quadratic
terms, we diagonalize $H_{S}$ as 
\begin{eqnarray}
H_{S} &=&\sum_{{\bf k}}[\omega _{{\bf k}}(\psi _{{\bf k}}^{\dag }\psi
_{{\bf k}}+\chi _{{\bf k}}^{\dag }\chi _{{\bf k}} +1) -12W]\ .
\end{eqnarray}
Here $\psi _{{\bf {k}}}$ and $\chi _{{\bf k}}$ are the quasiparticle
operators of the spin wave excitations with ${\bf k}$ the wave vectors
of one sublattice.  The quasiparticle spectrum is given by 
$\omega_{{\bf k}}=\sqrt{(W+P_{{\bf k}}^{-})^{2}-(P_{{\bf k}}^{+})^{2}}$, 
with $P_{{\bf k}}^{\mp }=2S\sum_{\alpha}
\tilde{J}_{\alpha }\cos k_{\alpha}\Theta (\mp \tilde{J}_{\alpha })$, and 
$W=2S\sum_{\alpha }|\tilde{J}_{\alpha }|$, in which 
$\Theta $ is the unit step function.

The orbital Hamiltonian $H_O$ looks
quite like $H_S$, where the orbital operator may be regarded as an isospin
operator. But  the absence of the SU(2) symmetry in $H_O$ and the abnormal
algebra of orbital operators make the orbital operators quite
different from the spin operators.
For example, orbital F-type arrangement is not an eigenstate of $H_O$,
and in case of orbital AF configuration,
on orbital sublattice $\bar{A}$ or $\bar{B}$ there
are only several preferable orbital alignments at which the
ground-state energy of the system reaches it minimum, unlike in a 
AF spin system where all the spin orientations on a sublattice are
energy-degenerate. 
In this case, the orbital state at site $i$ can be
generally expressed as $|i\rangle =\cos(\theta_{\sigma}/2)|z\rangle
+\sin(\theta_{\sigma}/2)|\bar{z}%
\rangle $ with $\sigma=+$ for $i\in \bar{A}$ and $-$ for $\bar{B}$,
respectively. From the symmetry of $h_x=h_y$ and relation
$m_{ix} + m_{iy} + m_{iz}=0$, the second term on the right-hand side
of Eq.\ (5) can be rewritten in a
more intuitive form 
\begin{equation}
H_z=-\varepsilon_z\sum_im_i^z\ ,
\end{equation}
with $\varepsilon_z=h_z-h_x$. This anisotropic Hamiltonian arises from 
anisotropy of electronic hopping integrals in orbital space as well as 
unusual algebra of orbital operators.
Both $u_{\alpha}$ and $h_{\alpha}$ are anisotropic and depend on the spin
configurations along the $\alpha$ direction, as shown in their expressions
below Eq.\ (5). Since $J_1-J_2+J_3$ is always positive,\cite{strength}
we have $u_z<u_x$ and $\varepsilon_z >0$  for the A-type AF spin
configuration; $u_z>u_x$ and $\varepsilon_z <0$ for the C-type AF one;
and $u_z=u_x$ and $\varepsilon_z =0$ for the ferromagnetic (F) one. 
The static JT distortions ${\bf Q}_{i}$ are
approximately treated as classical variables and assumed to be different
in the two sublattices, i.e., $%
Q_{i}\equiv Q_{\sigma}$ and $ \phi_{i}\equiv\phi_{\sigma}$ with
$\sigma=+$ $(-)$
for $i\in \bar{A}$ $(\bar{B})$. From x-ray diffraction experiments, it has
been confirmed that the MnO$_6$ octahedron is elongated along the $x$ or
$y$ direction, and the octahedrons are alternatively aligned in the 
$x$-$y$ plane, \cite{matsumoto} which corresponds to $\phi_{+}=2\pi/3$
and $\phi_{-}=-2\pi/3$ in the present formula.
Similar to the treatment of the spin degrees of freedom, we perform
the HP transformation for localized orbital operators. \cite{brink}
To the lowest order, $H_O$ can be diagonalized as
\begin{equation}
H_O=\sum_{{\bf k}\sigma}\varepsilon_{{\bf k}\sigma}\xi_{{\bf k}\sigma}^{\dag}
\xi_{{\bf k}\sigma}
   +\frac{1}{2}\sum_{{\bf k}\sigma}(\varepsilon_{{\bf k}\sigma}
   -P_{\sigma})+E_{C}\ .  \label{hp}
\end{equation}
Here $\xi_{{\bf k}\sigma}^{\dag}$ and $\xi_{{\bf k}\sigma}$ are the
quasiparticle operators of the orbital excitations, 
the second term stands for the quantum fluctuation energy where
\begin{eqnarray}
P_{\sigma}&=&-\sum_{\alpha}
4u_{\alpha}\cos\theta_{+}^{\alpha}\cos\theta_{-}^{\alpha}
+2\varepsilon_z\cos\theta_{\sigma}  \nonumber \\
&+&\frac{2g^2}{K}
\cos^2(\theta_{\sigma}-\phi_{\sigma})\ .  \nonumber
\end{eqnarray}
with $\theta^{\alpha}_{\sigma}=\theta_{\sigma}-\varphi_{\alpha}$,
and $E_C$ is the classical grand-state energy. The expression for 
$E_C$ depends on the orbital configuration. For both G- and C-type AF
configurations, it is given by  
$$
E_C/N = \sum_{\alpha}u_{\alpha}\cos\theta_{+}^{\alpha}
\cos\theta_{-}^{\alpha}-\frac{1}{2}\sum_{\sigma}[\varepsilon_z
\cos\theta_{\sigma}+
gQ_{\sigma}\cos(\theta_{\sigma}-\phi_{\sigma})-\frac{K}{2}Q^2_{\sigma}],
$$
with $N$ the number of the sites.
In principle, $\theta _{\sigma }$ and $Q_{\sigma }$ in Eq.(\ref{hp})
should be determined by minimizing the total ground state energy of the
system. In the present case, the quantum fluctuations in $H_{S}$ and
$H_{O}$ are small, and so the ground-state energy can be approximately
replaced by $E_{C}$.  It is found that besides the same
ground-state energy $E_C$, there is the same excitation spectrum 
for the C- and G-type AF orbital configurations, yielding
\begin{eqnarray}
\varepsilon_{{\bf k}\sigma}= \sqrt{\frac{1}{2}\{P^2_++P^2_-+\sigma[%
(P_+^2-P_-^2)^2+16P_+P_-C^2_{{\bf k}}]^{1/2}\}}\ , \nonumber
\end{eqnarray}
where
$C_{{\bf k}}=\sum_{\alpha}2u_{\alpha}\sin\theta_{+}^{\alpha}\sin%
\theta_{-}^{\alpha} \cos k_{\alpha}$.
This degeneracy of C- and G-type AF orbital configurations
agrees to Mizokawa and Fujimori's result. \cite{mizokawa}
Independent of the magnetic structure, such a degeneracy suggests the
possibility of a mixed C- and G-type AF orbital configuration in the
system, i.e, neighboring orbital states along the $z$ direction may be
either ``parallel" or ``antiparallel". In the absence of the
Coulomb interactions,
a C-type AF orbital structure may have lower energy.
\cite{hotta}

The JT coupling plays an important role in determining the orbital
ordering. In the absence of the JT coupling and in the small limit of
$\varepsilon _{z}$, the $e_{g}$ electrons may occupy two ``antiparallel"
states in the two sublattices:  
$(|z\rangle \pm |\bar{z}\rangle )/\sqrt{2}$ 
($\theta _{+}=-\theta _{-}=\pi /2$) for $u_{z}<u_{x}$;
$|z\rangle$ and $|\bar{z}\rangle $ 
($\theta _{+}=0,\theta _{-}=-\pi $) for $u_{z}>u_{x}$. 
Such symmetric "antiparallel" states will be broken by the uniform crystal
field appeared in Eq.\ (7). Furthermore, the JT distortions also lead to
an effective anisotropic crystal field acting on the two sublattices.
To distinguish it from the uniform crystal field $\varepsilon _{z}$, we
call it the JT field.  The JT field, whose strength increases with the
coupling constant $g$, tends
to align the orbital states in the two sublattices towards $|y\rangle $
($\theta_{+}=2\pi /3$) and $|x\rangle $ ($\theta _{-}=-2\pi /3$), 
respectively. 

The orbital ordering is described by the average value of operators
$m_i^{\alpha}$. From the orbital spectrum, it can be shown that
\begin{eqnarray}
\langle m^{\alpha}_{\sigma}\rangle =M_{\sigma}\cos\theta_{\sigma}^{\alpha}\ , 
\label{malpha}
\end{eqnarray}
with $\sigma=+$ $(-)$ for $i\in 
\bar{A}$ $(\bar{B})$, where 
\begin{eqnarray}
M_{\sigma}&=& 1-\sum_{\sigma^{\prime}}\int\frac{d^3k}{(2\pi)^3} 
\frac{2P_{%
\bar{\sigma}}C^2_{{\bf k}}}{\varepsilon_{{\bf k}\sigma^{\prime}}
[4P_+P_-C^2_{{\bf k}}+(P^2_{\sigma}-\varepsilon^2_{{\bf k}%
\sigma^{\prime}})^2]} 
\nonumber  \\
&\ &\times \left( \frac{2(P^2_{\sigma}+\varepsilon^2_{{\bf k}%
\sigma^{\prime}})} {e^{\beta\varepsilon_{{\bf k}\sigma^{\prime}}}-1}%
+(P_{\sigma}-
\varepsilon_{{\bf k}\sigma'})^2\right) \ .  \label{msigma}
\end{eqnarray}
The second term on the right-hand side of Eq.\ (10) comes from the 
quantum and thermal fluctuations. To keep a good approximation, this term
must be small at low temperatures.

We now discuss the ground state of the system. First, it is impossible
to realize an isotropic orbital ordering. Since $m_i^x+m_i^y+m_i^z=0$, if
$\langle m_i^x\rangle =\langle m_i^y\rangle =\langle m_i^z\rangle$, they
must be equal to zero and there is no any orbital ordering. 
From Eq.\ (4), it then follows that the anisotropy in $\langle
m_i^{\alpha}\rangle$ will lead to anisotropic $\tilde{J}_{\alpha}$.
At zero temperature, $M_{\sigma}= 1$ and
$\langle m^{\alpha}_{\sigma}\rangle=\cos\theta_{\sigma}^{\alpha}$
if the quantum fluctuation in Eq.(\ref{malpha}) is neglected. Taking
into account the symmetry requirement of
$\langle m_i^x + m_j^x\rangle =\langle m_i^y + m_j^y \rangle $,
we get two possible relations: (I) $\theta_{+}+\theta_{-}=0$ or (II)
$\theta_{+}-\theta_{-}=\pi$.  As the quantum fluctuation is taken into
account, relation (I) keeps unchanged, while relation (II) is
satisfied only approximately.
In both cases, we have $\tilde{J}_x=\tilde{J}_y\neq \tilde{J}_z$ from
Eq.\ (4), provided the small quantum fluctuations are neglected. 
Since the magnetic structure at zero temperature is determined by the
sign of $\tilde{J}_{\alpha}$, the same sign of $\tilde{J}_{\alpha}$,
regardless of anisotropic magnitude of them, will lead to a F or G-type
AF spin configuration, while different signs of $\tilde{J}_{x}$ and
$\tilde{J}_{z}$ will result in a A- or C-type spin configuration.
 Our calculations show that the ground-state magnetic structure is very
sensitive to the on-site Coulomb interactions.
Even though the magnetic superexchange $\tilde{J}_{AF}$ is fixed and 
the JT coupling is absent $(g=0)$, an evolution of spin configuration
in the order of $F\rightarrow A\rightarrow C \rightarrow G$
can be obtained with increasing the Coulomb interactions, 
as shown in Fig.\ 1(a).
It is found that spin configurations  A and G satisfy relation (I), and
spin configuration C satisfies relation (II). Figure 1(b) shows that 
an increasing JT coupling narrows gradually the C-type AF region.
This is because the JT coupling tends to align the orbital states along
$|x\rangle $ and $|y\rangle $,  and so raises the effective ferromagnetic
coupling in the $x$-$y$ plane and the AF coupling in the $z$ direction,
making the C-type AF spin configuration unstable.

We next discuss the orbital excitation spectra. Owing to the absence of
SU(2) symmetry in the orbital Hamiltonian, an orbital excitation spectrum
usually has an energy gap. For A-, C- and G-type AF  
spin configurations, there is always an energy gap in the orbital
spectrum, regardless whether or
not the JT coupling is taken into account (not shown here).
However, if the JT coupling is absent, gapless
orbital spectra may appear for the F spin configuration,
Furthermore, if relation (II)
is satisfied, the orbital spectrum has a two-dimensional form:
$\varepsilon_{{\bf k}\sigma}=6u_x\sqrt{1+\sigma(\cos k_x+\cos k_y)/2}$.
For such a two-dimensional spectrum, quantum and thermal fluctuations,
characterized by the second term of $M_{\sigma}$ in Eq.(\ref{msigma}),
will completely destroy long-range orbital ordering at finite
temperatures,\cite{hohenberg} resulting in an orbital-liquid state similar
to that obtained by Ishihara {\it et al}. \cite{ishihara2} 
The orbital excitation gap can be widened by the JT field acting on the
orbital states. It is very similar to an anisotropic magnetic
crystal field on the spin states in an AF Heisenberg Hamiltonian. Quantum
fluctuations are greatly suppressed by this JT field, making the orbital
ordering stable.

At finite temperatures, $\langle S^z_i\rangle $ and $M_{\sigma}$ in
Eq.\ (9) serve as the spin and orbital order parameters, respectively.
Both of them decrease with increasing the temperature, and $\langle
S^z_i\rangle $ ($M_{\sigma}$) vanishes as the temperature is increased
beyond the critical temperature $T_N $ ($T_O$). One may evaluate
$\langle S^z_i\rangle $ and $M_{\sigma}$ from a self-consistent equation
for $\langle S^z_i\rangle $ and Eq.\ (10). In our calculation, parameters
$J_1$, $J_2$ and $J_3$ are taken from the Racah parameters \cite{racah}
and $t=0.41$ eV.\cite{feiner} The system is found to have an A-type AF
spin configuration at low temperatures. In Fig.\ 2 we plot the variation
of $T_N$ and $T_O$ as functions of the strength of the JT coupling. Both
$T_N$ and $T_O$ increase with the JT coupling, but there are different
physical origins. The increase of $T_N$ is attributed to an enhancement
of the effective magnetic coupling $\tilde{J}_x$ and $\tilde{J}_y$
caused by the JT field. On the other hand, the increase of $T_O$ stems
from the fact that a stronger JT field will widen the energy gap of
the orbital excitation spectrum, and so a higher temperature is required
to excite orbital quasiparticles to break the long-range orbital ordering.
According to experimental data and theoretical analysis, $g$ is of the
same order of magnitude as $t$ and $K$ is greater than $g$ by a factor of
ten to hundred, \cite{okimoto,millis2} so that $g^2/K$ is the order of
$0.01t\sim 0.1t$. According to Fig.\ 2, to fit with $T_O=780 K$ measured
by the experiment, the strength of JT coupling should be $g^2/K=0.045t$,
at which the calculated $T_N=146 K$ is very close to the experimental
value of $T_N=140 K$.
The present calculation may overestimate the critical temperatures due to
neglecting the frequency-softened effect for the excitation spectrum at high
temperatures, and so the required strength of JT coupling may be greater
than the evaluated magnitude.

This work was supported by a CRCG grant at the University of Hong Kong.

\begin{figure}
\caption{Phase diagram at zero temperature in the absence (a) and 
presence (b) of the JT field. The parameters used are
$\tilde{J}_{AF}=0.006$
and $J_H=4/3$  with $t$ the unit of energy. The relation
$U=U^{\prime}+2J$
has been used \cite{strength}
and $U=20$ fixed in (b).}

\caption{Critical temperatures $T_N$ and $T_O$ as functions of $g^2/Kt$. }

\end{figure}


\begin{references}

\bibitem{goodenough}  J. B. Goodenough, Phys. Rev. {\bf 100}, 564 (1955).

\bibitem{murakami}  Y. Murakami {\it et al}., Phys. Rev. Lett. {\bf 81}, 582
(1998).

\bibitem{kugel}  K. I. Kugel and D. I. Khomskii, ZhETE {\bf 64}, 1429 (1973).

\bibitem{ishihara1} S. Ishihara, J. Inoue and S. Maekawa, Physica C {\bf 263},
130 (1996).

\bibitem{moreo} A. Moreo, S. Yunoki and E. Dagotto, Science {\bf 283}, 2034 (1999).


\bibitem{shiina} R. Shiina, T. Nishitani and H. Shiba, J. Phys. Soc. Jpn.
{\bf 66}, 3159 (1997).

\bibitem{maezono} R. Maezono, S. Ishihara, and N. Nagaosa, Phys. Rev. B
{\bf 58}, 11583 (1998).

\bibitem{brink}  J. van den Brink, P. Horsch, F. Mack and A. M. Ole\'{s},
Phys. Rev. B {\bf 59} 6795 (1999).


\bibitem{hotta} T. Hotta, S. Yunoki, M. Mayr and E. Dagotto, 
            cond-mat/9907034 (1999).


\bibitem{millis1} A. J. Millis, Phys. Rev. B {\bf 55}, 6405 (1996).

\bibitem{shen}  S. Q. Shen and Z. D. Wang, Phys. Rev. B {\bf 59},3291 (1999);
cond-mat/9906126 (1999).

\bibitem{feiner}  L. F. Feiner and  A. M. Ole\'{s}, Phys. Rev. B {\bf 59}, 3295
(1999).

\bibitem{strength}
A projective perturbation approach yields 
$J_{1}=t^{2}/[8(U+3J_{H}/2)]$, $J_{2}=t^{2}/[10(U^{\prime
}+3J/2+3J_{H}/2)]$, and $J_{3}=t^{2}/[10(U^{\prime }-J/2)]$.
Here $t$ is the hopping integral, $U$ ($U^{\prime}$) is the intra-
(inter-) orbital Coulomb interaction, $J$ and $J_H$ are the Hund's
coupling between the $e_g$ electrons and between the $e_g$ and the
$t_{2g}$ electrons, respectively. 

\bibitem{kanamori}  J. Kanamori, J. Appl. Phys. {\bf 31}, 14S (1960).

\bibitem{matsumoto} G. Matsumoto , J. Phys. Soc. Jpn. {\bf 29}, 606 (1970).

\bibitem{mizokawa} T. Mizokawa and A. Fujimori, Phys. Rev. B {\bf 51}, 12880
(1995); {\it ibid} {\bf 54}, 5368 (1996).

\bibitem{hohenberg} P. C. Hohenberg, Phys. Rev. {\bf 158}, 153 (1967).

\bibitem{ishihara2}  S. Ishihara, M. Yamanaka and N. Nagaosa, Phys. Rev. B
{\bf 56}, 686 (1997).

\bibitem{racah}
$U$, $ U^{\prime}$ and $J$ are connected through
three Racah parameters by $U=A+4B+3C,$ $ U^{\prime}=A-B+C$ and 
$J=5B/2+C$. Taking $A\approx 4.7 eV$ [Ref.\ 12], $B=0.107$ and $C=0.477 eV$
[A. E. Bocquet {\it et al}., Phys. Rev. B {\bf 46}, 3771 (1992)],
one has $U\approx 6.6, U^{\prime}\approx 5.1$ and $J=0.745 eV$.
$J_H\approx 1.2 eV$ is estimated from La$_{1-x}$Sr$_x$MnO$_3$ at
$x$=0.175. [M. Imada {\it et al.}, Rev. of Mod. Phys., {\bf 70}, 1039
(1998)].

\bibitem{okimoto}  Y. Okimoto {\it et al}., Phys. Rev. Lett. {\bf 75}, 109 
(1995); Phys. Rev. B {\bf 55}, 4206 (1997).

\bibitem{millis2}  A. J. Millis, Phys. Rev. B {\bf 53}, 8434 (1996); Y.
Motome and M. Imada, J. Phys. Soc. Jpn. {\bf 68}, 16 (1999).

\end{references}
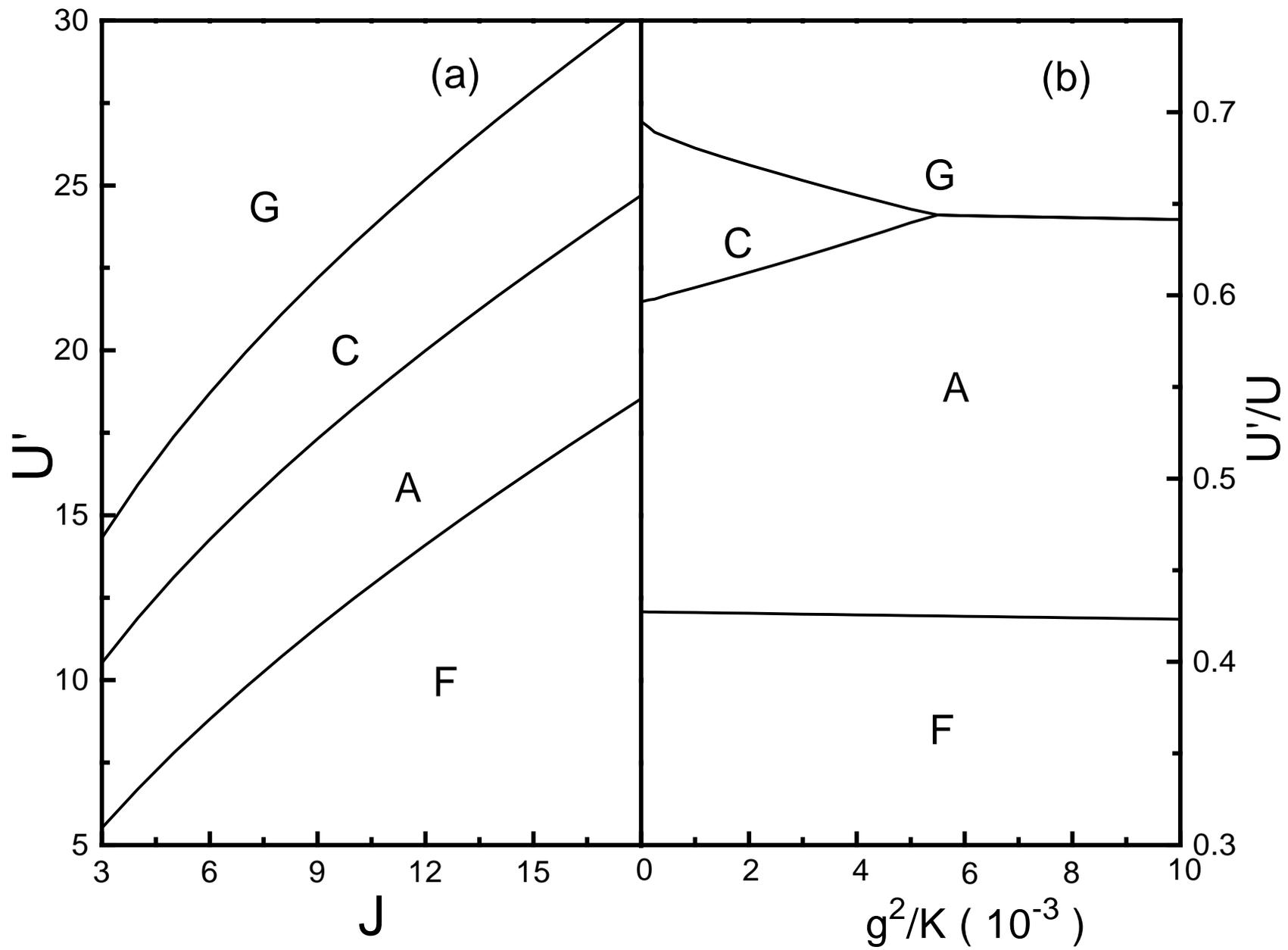
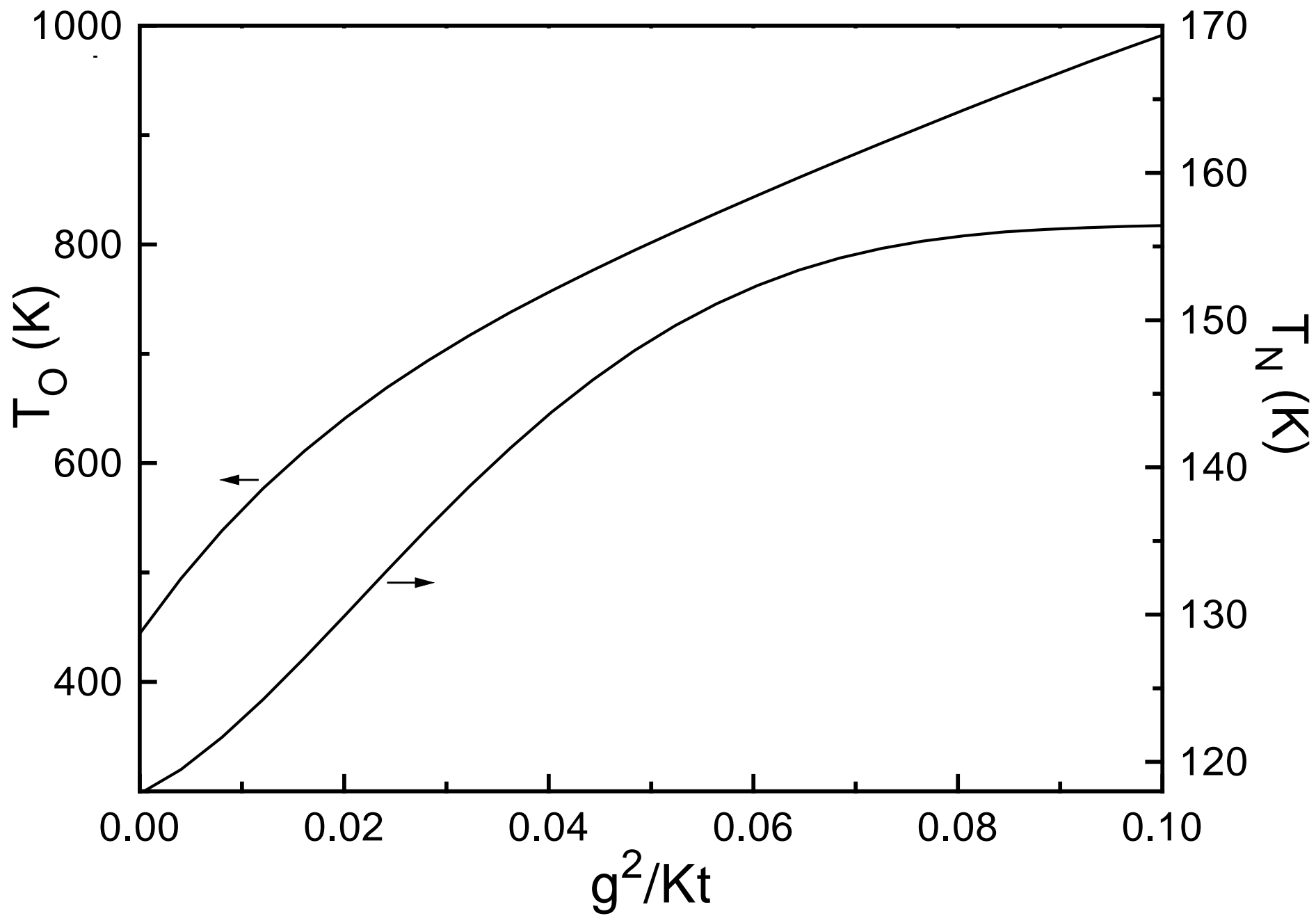
\end{document}